\documentclass[%
 reprint,
nofootinbib,
 amsmath,amssymb,
 aps,
]{revtex4-2}

\usepackage{graphicx}% Include figure files
\usepackage{xcolor}
\usepackage{dcolumn}% Align table columns on decimal point
\usepackage{bm}% bold math
\usepackage{hyperref}% add hypertext capabilities
\usepackage{cleveref}
\newcommand{\dd}{\text{d}}
\newcommand{\ii}{\text{i}}

\begin{document}

\title{Comments on the validity of the non-stationary Generalized Langevin Equation as a coarse-grained evolution equation for microscopic stochastic dynamics}% Force line breaks with \\
\thanks{This article may be downloaded for personal use only. Any other use requires prior permission of the author and AIP Publishing. This article appeared in \cite{doi:10.1063/5.0049693} and may be found at \url{https://doi.org/10.1063/5.0049693}.}%

\author{Fabian Glatzel}
\email{fabian.glatzel@physik.uni-freiburg.de}
\author{Tanja Schilling}%
\email{tanja.schilling@physik.uni-freiburg.de}
 
\affiliation{%
 Institute of Physics, University of Freiburg, Hermann-Herder-Str. 3, 79104 Freiburg, Germany
}%

\date{\today}% It is always \today, today,

\begin{abstract}
	We recently showed that the dynamics of coarse-grained observables in systems	out of thermal equilibrium are governed by the non-stationary generalized	Langevin equation [J.~Chem.~Phys.~{\bf 147}, 214110 (2017),	J.~Chem.~Phys.~{\bf 150}, 174118 (2019)]. The derivation we presented in these two articles was based on the assumption that the dynamics of the microscopic degrees of freedom was deterministic. Here we extend the discussion to	stochastic microscopic dynamics. The fact that the same form of the	non-stationary Generalized Langevin Equation as derived for the deterministic case also holds for stochastic processes implies that methods	designed to estimate the memory kernel, drift term and fluctuating force term	of this equation as well as methods designed to propagate it numerically, can	be applied to data obtained in molecular dynamics simulations that employ a	stochastic thermostat or barostat.
\end{abstract}

\maketitle

%\tableofcontents
\section{Introduction}

The concept of projection operators as a tool to reduce the dimension of physical systems goes back to the 1960s. Based on work by Nakajima \cite{nakajima1958}, Zwanzig showed how projection operators can be used to derive an equation of motion for almost arbitrary phase space observables \cite{PhysRev.124.983,doi:10.1063/1.1731409}. Five years later, Mori derived an analogous result \cite{10.1143/PTP.33.423} employing a different kind of projection operator. In more recent works of e.g.~Chorin et al. \cite{PMID:10737778,10.2140/camcos.2006.1.1}, the topic is addressed in a more general mathematical setting abstracting from the context of physical observables and phase space.

Grabert showed in the 1970s how the projection operator technique can be extended to non-stationary processes by using time-dependent projection operators \cite{grabert1977microdynamics,grabert2006projection}. However, Graberts approach was very general and did not include any explicit generalization of either Zwanzig's or Mori's projection operator to non-equilibrium dynamics. (Notably, a derivation of a non-stationary equation of motion using a time-dependent generalization of Mori's projection operator formalism was presented by Nordholm in his PhD thesis as early as 1972 \cite{nordholm1972}, but unfortunately this was not taken up by the community as it was not published elsewhere.)

Concrete applications of time-dependent projection operators to model colloidal suspensions and undercooled liquids were provided by Shea, Oppenheim and Latz in the 1990s \cite{shea1996fokker, shea1997fokker, shea1998fokker, latz2002non}. In recent work Meyer et al.~\cite{doi:10.1063/1.5006980, doi:10.1063/1.5090450} as well as te Vrugt et al.~\cite{Vrugt2019, Vrugt2020} used time-dependent projection operators similar to Mori's 
to derive a general equation of motion for coarse-grained variables 
in non-equilibrium systems, including even systems under time-dependent external driving.
(A similar attempt was made by Kawai and Komatsuzaki via the Zwanzig projection operator, but it turned out to be more involved mathematically than the Mori approach\cite{Kawai11}.) Further, Meyer et 
al.~introduced a general and fast method to compute the memory kernels appearing in their non-stationary Generalized Langevin Equation (abbreviated as ``nsGLE'' and always refering to the version by Meyer et al.) from a set of time-resolved values of an observable for individual trajectories \cite{10.1002/adts.202000197}. As such data is often accessible through molecular dynamics simulations, the question arises naturally if one can apply the same formalism in the context of stochastic microscopic propagators (e.g.~dynamics generated using thermostats and barostats). This is not directly clear because the derivation of the nsGLE demands a deterministic Liouvillian.

Espa{\~n}ol and V{\'a}zquez have used a projection operator formalism of the Zwanzig-type in order to coarse-grain dynamics which are governed by the Fokker-Planck equation \cite{Espanol}.  Importantly, they observed that under conditions of time-scale separation, the coarse-grained dynamics is again governed by a 
Fokker-Planck equation. A similar but more explicit route has been taken in the work of Kranz et al.~\cite{PhysRevE.87.022207}. In both cases, the 
average over the stochastic degrees of freedom was taken before the projector was applied. Here we take a different route in order to obtain an equation of motion that treats as independent, trajectories which do not differ in their initial physical configuration but only in the explicit stochastic contribution. To illustrate that the framework 
can in fact be extended to stochastic processes in such a way, we will start with a brief recapitulation of the derivation of the nsGLE in \cref{sec:recap} before discussing the case of stochastic dynamics in \cref{sec:stochasticity}.

\section{Recapitulation and Notation}\label{sec:recap}
Given an initial phase space distribution $\rho(\Gamma,0)$, that is neither necessarily stationary nor, in particular, the equilibrium distribution, and a (time-dependent) Liouvillian $\mathcal{L}_t$, the phase space distribution is determined for any future time by the Liouville equation. Here, $\Gamma$ denotes the collective set of phase space coordinates. If one intends to apply a time-dependent projection operator to a time-dependent Liouvillian, it can be useful to switch from the usual phase space representation to an ``augmented'' phase space that includes one additional coordinate for the system time $\Gamma':=(\Gamma,\tau)$\cite{doi:10.1063/1.5090450}. Then a Liouvillian $\mathcal{L}'$ in the augmented phase space can be defined generating the original dynamics without an explicit time-dependence ($\mathcal{L}'$ does not depend on $t$) and projections can be carried out on this Liouvillian. Unless stated otherwise, the 
following calculations in this section are carried out in the augmented phase space and the prime in the notation is dropped for now.

One can write the time evolution of an observable $\mathbb{A}(\Gamma)$, that is a function of the augmented phase space coordinates, as
\begin{align}\label{eq:timeEvolutionObservable}
	A_t&=\mathbb{A}(\Gamma_t) = \exp(\ii\mathcal{L}t)\mathbb{A}(\Gamma)\big|_{\Gamma_{0}}.
\end{align}
Here, $\Gamma_0$ denotes the initial point in the augmented phase space of one trajectory, $\Gamma_t$ is the point in the augmented phase space reached by the same trajectory at time $t$, and $A_t$ is the value of the observable for a specific trajectory as a function solely of time. We implicitly assumed that the dynamics can be described by analytic functions on the whole observation interval. To allow for easy readability, we will omit spelling out the dependencies on the phase space coordinates of $\mathbb{A}(\Gamma)$ and the insertion of the initial point in phase space $\Gamma_0$ from now on.

By taking the time derivative of \cref{eq:timeEvolutionObservable} and using Graberts approach \cite{grabert1977microdynamics,grabert2006projection} of applying time-dependent projection operators one obtains
\begin{subequations}
	\begin{align}
		\frac{\dd A_t}{\dd t} &= \exp(\ii\mathcal{L}t)\ii\mathcal{L}\mathbb{A}\\
		&= \exp(\ii\mathcal{L}t)\mathcal{P}_t\ii\mathcal{L}\mathbb{A}\nonumber\\
		&\phantom{=}+\int\limits_0^t\dd\tau\,\exp(\ii\mathcal{L}\tau)\mathcal{P}_\tau (\ii\mathcal{L}-\dot{\mathcal{P}}_\tau)\mathcal{Q}_\tau\mathcal{G}_{\tau,t}\ii\mathcal{L}\mathbb{A}\nonumber\\
		&\phantom{=}+\mathcal{Q}_0\mathcal{G}_{0,t}\ii\mathcal{L}\mathbb{A}.\label{eq:generalProjectionOperatorIdentity}
	\end{align}
\end{subequations}
Here, $\mathcal{P}_t$ is a time-dependent projection operator, $\mathcal{Q}_t:=1-\mathcal{P}_t$ is its orthogonal complement and
\begin{align}
	\mathcal{G}_{s,t} &:=\exp_-\left(\int\limits_{s}^t\dd\tau\,\ii\mathcal{L}\mathcal{Q}_\tau\right),
\end{align}
where $\exp_-$ denotes a negatively time-ordered exponential function. Next, we introduce a time-dependent product on the set of phase space observables by
\begin{align}
	(\mathbb{X},\mathbb{Y})_t &= \int\dd\Gamma \rho(\Gamma,0)\left(\exp(\ii\mathcal{L}t)\mathbb{X}\right)\left(\exp(\ii\mathcal{L}t)\mathbb{Y}\right),
\end{align}
where $\rho(\Gamma,0)$ is the initial probability density in the augmented phase space obtained by multiplying the probability density of the initial ensemble with the term $\delta(\tau)$ syncing the observation time and the augmented time coordinate. One can define a specific projection operator by
\begin{align}
	\mathcal{P}_t \mathbb{X}:= \frac{\left(\mathbb{X},\mathbb{A}\right)_t}{\left(\mathbb{A},\mathbb{A}\right)_t}\mathbb{A},
\end{align}
where $\mathbb{A}$ is a specific observable of interest and $\mathbb{X}$ is any phase space function.

Inserting this explicit projection operator, \cref{eq:generalProjectionOperatorIdentity} can be rewritten as
\begin{align}
	\frac{\dd A_t}{\dd t} &= \omega(t)A_t+\int\limits_0^t\dd\tau\,K(t,\tau)A_\tau+\eta_t.
\end{align}
Here, the quantities $\omega(t)$, $K(\tau,t)$ and $\eta_{t}$ are defined by
\begin{subequations}
	\begin{align}
		\omega(t) &:= \frac{\left(\ii\mathcal{L}\mathbb{A},\mathbb{A}\right)_t}{(\mathbb{A},\mathbb{A})_t},\\
		K(t,\tau) &:= \frac{\left(\left(\ii\mathcal{L}-\dot{\mathcal{P}}_\tau\right)\mathcal{Q}_\tau\mathcal{G}_{\tau,t}\ii\mathcal{L}\mathbb{A},\mathbb{A}\right)_\tau}{(\mathbb{A},\mathbb{A})_\tau},\\
		\eta_{t} &:= \mathcal{Q}_0\mathcal{G}_{0,t}\ii\mathcal{L}\mathbb{A}
	\end{align}
\end{subequations}
Note that the functions $\omega(t)$ and $K(t,\tau)$ are the same for different trajectories, if their initial conditions are drawn from one given initial probability density (i.e.~for trajectories drawn out of one given 
non-equilibrium ensemble or ``swarm'' of trajectories). Hence, we denote their time dependencies using parenthesis, whereas quantities with time as a subscript, such as $A_t$ and $\eta_t$, do depend on the individual trajectory. $K(t,\tau)$ is the so called memory kernel describing non-local 
(in time) contributions for the equation of motion of $A_t$. It can be shown that these quantities fulfill a relation that is similar in structure 
to a fluctuation-dissipation theorem \cite{doi:10.1063/1.5006980}, namely
\begin{align}\label{eq:genalizedFDT}
	\langle\eta_{t_1}\eta_{t_2}\rangle &= -K(t_1,t_2)\,(\mathbb{A},\mathbb{A})_{t_2}.
\end{align}

\section{Including Stochasticity}\label{sec:stochasticity}

The derivation of the nsGLE presented in Refs.~\cite{doi:10.1063/1.5006980,doi:10.1063/1.5090450} requires taking time-derivatives of the observable. The derivation can therefore not be directly applied to stochastic dynamics. However, the idea of the augmented phase space can be used to to 
tackle this problem.

\begin{figure}
	\centering
	\includegraphics[scale=1]{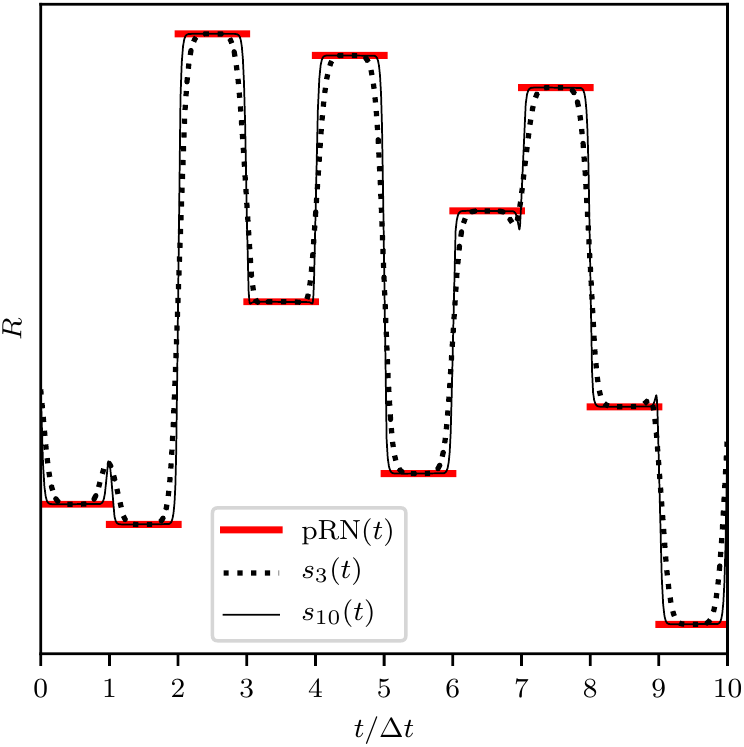}
	\caption{Sketch of noise generated by a pseudorandom number generator and its smoothed version.}
	\label{fig:rng2}
\end{figure}

We construct an augmented phase space such that we can keep track of the random numbers generated along a single trajectory, i.e.~of the values taken by the noise function at each time step. Here, we assume that there is only a finite number of random numbers per trajectory, which is the case for computer simulations, in which time is discretized. If we refer to the set of random numbers per trajectory by $R$, then the augmented phase 
space is given by $(\Gamma'):=(\Gamma,\tau,R)$, where we have already included the time-coordinate $\tau$ for convenience. Further, we note that 
the Liouvillian does not affect the random variables ($\ii\mathcal{L}R_i=0\quad\forall R_i\in R$) and, thus, they may be regarded as integrals of motion. The values of the random variable taken along a single trajectory, when interpreted as a function of simulation time, may look like the bold red line in \cref{fig:rng2}. We labeled the line by pRN for ``pseudo random number'' as this is the case in a typical computer simulation, but the arguments hold equally well for numbers generated by a truly stochastic process.

Note that the pRN data has discontinuous jumps from one time step to another and, hence, naive coupling of physical degrees of freedom to the bold 
red pRN curve would introduce non-analytic behavior. However, we argue in 
the following that one can interpret this data as the limit of a series of analytic functions. Let us take a trajectory with $M$ time steps in total and random numbers given by $R_1,\cdots,R_M$. To keep the notation simple, we will assume that $\Delta t=1$ and that the jumps of the pRN data occur in the middle between consecutive time steps. Then, the value of the pRN as function of time can be expressed through
\begin{align}
\text{pRN}(t) &= \sum\limits_{i=1}^M R_i\, \theta\left(1-2\left|i-t\right|\right).\label{eq:randomTraj}
\end{align}
Here, $\theta(t)$ denotes the Heavyside step function. In contrast to $\text{pRN}(t)$, eq.\ref{eq:randomTraj}, the function defined through
\begin{align}
s_n(t) &= \sum\limits_{i=1}^M R_i\, \exp\left(-(2(i-t))^{2n}\right)\label{eq:smoothRandomTraj}
\end{align}
with $n\in\mathbb{N}$ is analytic and convergent on $\mathbb{R}$. (Note that we can assume that the Liouvillian couples some physical degree of freedom to the random variables as given in \cref{eq:smoothRandomTraj}, because the augmentation of the phase space includes a time coordinate.) \Cref{fig:rng2} shows two exemplary curves of $s_n(t)$ for two different values of $n$. $s_n(t)$ converges pointwise to $\text{pRN}(t)$ in the limit $n\to\infty$ if one neglects the null set of points where the jumps for the $\text{pRN}(t)$ occur. Molecular dynamics simulations are aimed at a numerical integration and, hence, one needs to check that also the 
limit of the integrals over \cref{eq:smoothRandomTraj} converges to the integral over \cref{eq:randomTraj}. Here, pointwise convergence is not a sufficient condition. However, if the values of the pRN are bounded, $s_n(t)$ is uniformly integrable $\forall n\in\mathbb{N}$ and the convergence of the integrals is given as well.

Finally, we need to check that also the value of the observable along the 
trajectory produced using $s_n(t)$ converges to the one obtained by using 
$\text{pRN}(t)$. As our observable is an analytic function of the phase space coordinates, it suffices to check that the distance in phase space between trajectories produced using $s_n(t)$ and $\text{pRN(t)}$ vanishes in the limit $n\to\infty$. However, this point needs some additional consideration.

The equations of motion for the two cases, i.e.~dynamics with the pseudo random numbers generating the ``noise'' and dynamics with the analytic term \ref{eq:smoothRandomTraj}, will generate different trajectories. In chaotic systems the distance between the trajectories is expected to grow in time. More precisely, if one defines a metric of phase space to quantify the deviation of two states, Oseledets's theorem states that this deviation grows (or shrinks) asymptotically exponentially (described by its Lyapunov exponents) for non-integrable systems \cite{Oseledets}. 

Now assume that one intends to analyze molecular dynamics trajectories of 
some finite duration. The deviation between two final states after the complete simulation-time due to some ``small perturbation'' at the beginning will always remain finite -- even if it is usually very large -- and hence it can be scaled down by weakening the initial perturbation, which in 
turn can be achieved by increasing $n$. Using the Lyapunov exponents of the given system and demanding that the trajectories deviate less than some desired value (e.g.~a value on the order of numerical precision), it is 
straight forward to find some threshold for the allowed deviation after a 
single integration time-step. However small this threshold may be, naturally one can choose a value of $n$ large enough such that the two dynamics, the one with $\text{pRN}(t)$ and one with $s_n(t)$ data, deviate less than that after a single integration time-step. Further, assuming a Liouvillian generating analytic dynamics, not only the deviation in the long-time limit but also short-time fluctuations between the two dynamics can be 
reduced to an arbitrarily small but nonzero value.

Formally, this can be expressed in the following way:\\
Let $\Delta\Gamma(t)$ be the distance between two initially close trajectories in the physical degrees of freedom of phase space (e.g. $\Delta\Gamma(t)=\left(\Delta q_1(t),\cdots,\Delta q_N(t),\Delta p_1(t),\cdots,\Delta p_N(t)\right)^T$ with the generalized coordinates $q_i(t)$ and momenta $p_i(t)$). Further, with a meaningful norm $\|\Delta\Gamma(t)\|$, e.g. $\|\Delta\Gamma(t)\|=\sqrt{a_1\Delta q_1(t)^2+\cdots+a_{2N}\Delta p_{N}(t)^2}$ where the $a_i$ have a strictly positive value and cancel the units of the corresponding factors, one gets $\|\Delta\Gamma(t)\|\propto\exp(\lambda t)$ in the long time limit. Here, $\lambda$ is the largest Lyapunov exponent. By demanding that at the end of the simulation time $t=T$ 
the norm of the deviation takes some arbitrarily small but finite value $\|\Delta\Gamma(T)\|$, one can now calculate a threshold for the deviation 
after a single simulation-step, namely
\begin{align}
\|\Delta\Gamma(\Delta t)\| &= \exp(\lambda (\Delta t-T))\,\|\Delta\Gamma(T)\|.
\end{align}
This value may become absurdly small but will always remain finite. Next, 
we assume that the initial separation is small ($\Delta\Gamma \to \dd\Gamma$) and that the flow field $\dot{\Gamma}(\Gamma)$ is bounded. (Note that $\dot{\Gamma}(\Gamma)$ as a function of the complete phase space will in most cases be unbounded. However, any finite subspace of finite trajectories will lie in some compact subset of phase space. By demanding $\dot{\Gamma}(\Gamma)$ to be analytic it must also be continuous and hence $\dot{\Gamma}(\Gamma)$ is bounded in this subspace which suffices for the following considerations.) Then, we can write
\begin{align}
\|\dd\dot{\Gamma}(t)\| &= \left\|\frac{\dd\dot{\Gamma}(\Gamma(t))}{\dd\Gamma(t)}\,\dd\Gamma(t)\right\|\leq c\,\|\dd\Gamma(t)\|
\end{align}
with some constant $c$. Hence, it is clear that also the short time fluctuations become arbitrarily small as the initial separation diminishes.

Thus, by reinterpreting the noise as a function obtained as a limit of analytic functions, one can analyze data generated with stochastic propagators by applying the methods provided by the \mbox{nsGLE}. Note that we made no assumption on how the random variables $R$ for each trajectory are obtained. Hence, the above reasoning holds for both pseudorandom numbers and truly random numbers. Further, we note that the probability distribution of the random numbers enters the initial phase space probability distribution $\rho'(\Gamma', 0)$. In the simplest case when the initial phase 
space probability distribution of the physical degrees of freedom and the 
random degrees of freedom are independent, which is by no means necessary 
nor demanded by the formalism of the nsGLE, one could write
\begin{align}
\rho'(\Gamma',0) &= \rho(\Gamma,0)\,\delta(\tau)\,p(R),
\end{align}
where $\rho(\Gamma,0)$ is the initial phase space density in the original 
(i.e.~not augmented) phase space, $\delta(\tau)$ is the contribution syncing the augmented phase space coordinate $\tau$ with the observation time $t$ (cf. \cite{doi:10.1063/1.5090450}), and $p(R)$ is the joint probability distribution of all degrees of freedom describing the random noise. Note, that in the more general case where the probability distribution of the initial physical degrees of freedom and the stochastic ones do not separate, every initial physical configuration can have its own probability distribution of the stochastic degrees of freedom allowing for stochastic contributions that  differ for different trajectories.

The arguments above handled the special case where the noise is a step function. This case is quite common in the context of computer simulations, 
but we will generalize the approach in the following to allow for more general types of stochastic processes. First, we introduce a new space, which will replace the augmented phase space. Given that the initial phase space of our system is of type $\mathbb{R}^N$, the new space will be of the type $\mathbb{R}^N\times\mathbb{R}\times F^N$, where $F$ is the space of power series converging in the whole interval of observation. Here, $\mathbb{R}^N$ contains again the physical degrees of freedom and the additional $\mathbb{R}$ contains, again, the degree of freedom for the trajectory time. Further, for every physical degree of freedom $q_i$ there is now 
an analytic function $f_i(x)\in F$ that will describe the realization of the stochastic properties. We will denote points in this space by $(\Gamma')=(\Gamma,\tau,f)$. The Liouvillian in this space can then be written as
\begin{align}
	\ii\mathcal{L}' &= \ii\mathcal{L}+\sum\limits_i\int\dd x\, f_i(x)\delta(x-\tau)\frac{\partial}{\partial q_i}+\frac{\partial}{\partial\tau},
\end{align}
where $\mathcal{L}$ is the Liouvillian of the original system, acting and 
depending only on $(\Gamma)$ and $q_i$ are the physical degrees of freedom. Hence, the action on the physical degrees of freedom is given by
\begin{align}
\ii\mathcal{L}'q_i &= \ii\mathcal{L}q_i + \int\dd x\,f_i(x)\delta(x-\tau) = \ii\mathcal{L}q_i + f_i(\tau),
\end{align}
whereas the $f_i(x)$ remain unchanged under the action of the Liouvillian 
($i\mathcal{L}'f_i(x) \equiv 0$).

If we consider an observable $\mathbb{A}'(\Gamma')$ which depends only on 
the physical degrees of freedom $(\Gamma)$ for which $\ii\mathcal{L}'q_i=\dd q_i/\dd t$, we can write
\begin{align}
\frac{\dd A_t}{\dd t} &= \exp(\ii\mathcal{L}'t)\ii\mathcal{L}'\mathbb{A}\big|_{\Gamma_{0}'}.
\end{align}
From this, we can derive the nsGLE analogously as before. However, for the application of the projection operators we need the probability density $\rho'(\Gamma',0)$ and we need to carry out integrals over all degrees 
of freedom of the augmented phase space. To circumvent the problem of needing a measure for this infinite dimensional space, we point out that if the analytic functions can be specified by a finite number of real control parameters $\mathbb{R}^M$, we could introduce a mapping $\mathbb{R}^M\to F^N$, define the probability density in these coordinates, and use a simple measure of the $\mathbb{R}^{N+1+M}$, e.g.~the Lebesgue measure.

An example for such a mapping could be a particle moving in three dimensions and getting random kicks. Assume that these kicks can be described by 
a force that is a Gaussian in time. So, every kick is determined by the time of occurrence (center of Gaussian), the width of the Gaussian, an amplitude, and a direction (e.g. specified by two real numbers). Hence, every individual kick can be specified by five real numbers.

Note that the arguments presented here apply to coarse-graining of explicitly time-dependent Liouvillians and to any kind of stochastic process which is bounded and either time-discrete, or if it is time-continuous, each realization of which can be approximated by an analytic function. This constitutes a generalization compared to previous work \cite{Espanol}, which applies to the case where the Liouvillian is not explicitly time-dependent and the stochastic microscopic process can be described by a Fokker-Planck equation.

\section{Conclusion}\label{sec:conclusion}
We have shown that the general framework of the non-stationary Generalized Langevin Equation can be applied to a wide range of processes with (pseudo)stochastic contributions. This includes particularly simulation data obtained using some kind of stochastic propagator, such as e.g.~a thermostat or barostat. The generalization was achieved by further abstraction from the usual phase space towards a more convenient space that includes degrees of freedom capturing the stochastic contributions.

\section{Acknowledgments}
The authors acknowledge funding by the Deutsche Forschungsgemeinschaft (DFG, German Research Foundation)—Project No.~430195928, and useful mathematical remarks from Fabian Coupette.

\section{Data Availability}
Data sharing is not applicable to this article as no new data were created or analyzed in this study.

\end{document}